# Surface modes and parity violation in Schwinger model on the lattice

I. Horváth[a*]

[a]SCRI, Florida State University, Tallahassee, FL 32306-4052, USA

The phase diagram of the Schwinger model on the lattice with Wilson fermions is investigated in the Hartree-Fock approximation. In case of single flavour (not directly amenable to simulation), the calculation indicates the existence of the parity violating phase at both weak and intermediate-to-strong couplings. Hartree-Fock vacuum sustains a nonzero electric field in this broken phase. The phase structure of the model with two flavours is also discussed.

## 1. INTRODUCTION

Understanding chiral symmetry on the lattice has become a rather popular subject during the last couple of years and some significant progress has been achieved (see for example the review talk by Y. Shamir in this volume). With this wave of new activity some old problems in standard formulations are being revived, such as the question of chiral limit in vectorlike theories with Wilson fermions. In particular, one would like to relate the concepts of chiral symmetry breaking that we believe in in continuum, to the lattice concepts at nonzero lattice spacing where chiral symmetry is not present in the first place.

A possible scenario proposed some time ago by Aoki [1] ascribes the masslessness of pions that is generally claimed to be observed at critical hopping parameter in numerical simulations of QCD, to the spontaneous breakdown of parity and flavour occuring at this point. In particular, he assumes that upon crossing the critical line ($\kappa > \kappa_c(\beta)$), one enters a phase with $\langle i\bar\psi\gamma_5\tau_3\psi\rangle \neq 0$ (two flavours) or $\langle i\bar\psi\gamma_5\psi\rangle \neq 0$ (one flavour). While well tested at strong coupling, the above picture is far from being conclusively established at weak coupling, close to the continuum limit.

In Ref. [2] the mechanism for generating the parity violating phase in case of one flavour has been traced to the existence of the surface modes of Wilson fermions on a finite lattice with open boundaries. This is most easily seen in one spatial dimension: As hopping becomes critical, two surface modes of almost zero energy appear bound to the ends of the system and the left-right asymmetry is introduced as only one of them gets filled to form the Dirac sea. As a consequence it is assumed that, at least at weak coupling, the vacuum will sustain a nonzero electric field (with U(1) gauge group) at the supercritical hopping. These ideas were further developed in an inspirative review by Creutz [3], concluding the qualitative behaviour of phase diagrams for different number of flavours and spatial dimensions.

It is with these concepts in mind that I investigate the global phase structure of the Schwinger model with Wilson fermions in this work. In other words, the idea is to examine the Aoki's scenario in this context (which is frequently considered in analogy to QCD in four dimensions), guided by the surface mode picture which is very valuable because of its qualitative predictive power. Aside from that, it is interesting to note that surprisingly little is known about the Schwinger model with Wilson fermions. This is especially true for the model with one flavour, where there is no standard way to simulate the system. This has recently been stressed by Gausterer and Lang [4] who studied the system at very strong coupling. The Schwinger model was also studied in Ref. [5]. In what follows, I will describe the main results of the Hartree-Fock (H-F) analysis of the phase diagram for one and two flavours. A detailed account and extension of this work is in preparation.

*Work supported by the DOE under Grant Nos. DE-FG05-85ER250000 and DEFG05-92ER40742.

## 2. ONE FLAVOUR

I will consider the lattice Hamiltonian

$$H = H_W - \frac{g^2}{4} \sum_{n,m} \rho_n \mid n - m \mid \rho_m. \qquad (1)$$

Here $H_W$ is the Hamiltonian of free Wilson fermions, including the axial (parity violating) mass term, namely

$$\begin{aligned} H_W &= K \sum_n \overline{\psi}_{n+1}(i\gamma_1 - r)\psi_n + h.c. \\ &+ \sum_n [M\overline{\psi}_n\psi_n + M_5 i\overline{\psi}_n\gamma_5\psi_n], \end{aligned} \qquad (2)$$

and the second term is the interaction. $K, r, M$ and $g$ are the hopping parameter, Wilson parameter, mass and gauge coupling respectively. The charge density operator is dedfined as $\rho_n = \psi_n^\dagger \psi_n - 1$. Restriction to the charge zero sector then requires the system to be half-filled. The fermionic operators $\psi_j$ are subject to canonical anticommutation relations and the boundary conditions on a finite lattice of size $L$ are opened. The above lattice theory represents a direct discretization of the continuum model in axial gauge.

H-F approximation to the vacuum of the theory amounts to finding the lowest energy state in the space of $L$-particle Slater determinants over all complete sets of normalized one-particle states $\{\phi_n^\alpha, \alpha = 1, \ldots, 2L\}$. This variational problem can be solved by subjecting these one-particle states to the H-F equations of the form

$$\sum_m \tilde{H}_{nm} \phi_m^\alpha = \epsilon^\alpha \phi_n^\alpha, \qquad (3)$$

with

$$\begin{aligned} \tilde{H}_{nm} &= K_{nm} + \frac{g^2}{2}\Big[ V_{nm}^D + V_{nm}^E \Big] \\ V_{nm}^D &= \delta_{nm} \sum_j \mid n - j \mid (1 - \sum_{\beta=1}^L \phi_j^{\beta\dagger}\phi_j^\beta) \\ V_{nm}^E &= \sum_{\beta=1}^L \phi_n^\beta \mid n - m \mid \phi_m^{\beta\dagger}. \end{aligned} \qquad (4)$$

Here $K_{nm}$ is the one-particle Hamiltonian, corresponding to $H_W$ and $V_{nm}^D, V_{nm}^E$ are the direct and the exchange potential respectively. The above equations have to be solved self-consistently. Let me also remark that all the discrete symmetries (including parity) of the full theory are preserved by its H-F approximation.

I have solved the H-F equations on finite lattices numerically. The details of the numerical procedure will be given elsewhere. In computations, I have set $K = 1$ and $r = 1/2$. With this normalization, the surface modes at zero coupling exist in the region $M < M_c(0) = 1$ [2]. In this single flavour case, I have mostly concentrated on the phase diagram in the $M - g^2$ plane. To be able to explicitly observe expectation values of parity-odd operators on finite lattices, I have set the axial mass to a very small value, $M_5 = 10^{-3}$.

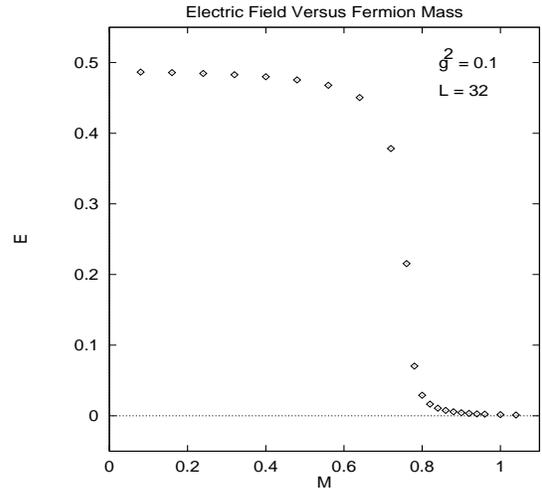

Figure 1. Vacuum expectation value of the electric field in the middle of the system as a function of fermion mass.

The representative example of the most relevant finding in this study is displayed in Fig.1. The vacuum expectation value of the electric field (in units of $g$) on a 32-site lattice is ploted as a function of the fermion mass at weak coupling. Note that for large values of $M$, electric field tends to zero as one would expect from the parity-conserving theory. However at small fermion masses the field acquires an expectation value and the two regions are separated by a rapid transition. This suggests the existence of the parity vi-olating phase transition and confirms the qualitative picture presented in Refs.[2,3] at the Hartree-Fock level. The physics of the broken phase corresponds to $\theta = \pi$ in continuum.

Similar behaviour is observed also at higher values of gauge coupling. The resulting positions of phase transitions observed on the lattice with 48 sites are ploted in Fig.2. Diamonds sample the critical line $M_c(g^2)$, with parity broken in the left region. The transition point $M_c(g^2)$ is defined here simply as the fermion mass, at which the expectation of electric field rises above the value one order of magnitude larger than the size of the parity-violating axial term, namely $10^{-2}$. The square marks $M_c(0)$ at infinite volume limit. The left vertical line represents $M_c(\infty)$ quoted in [4] for the model in standard Lagrangian formulation. By comparing to the results on smaller lattices, I expect the critical masses at nonzero couplings to be increased by few parts in hundred in the infinite volume limit.

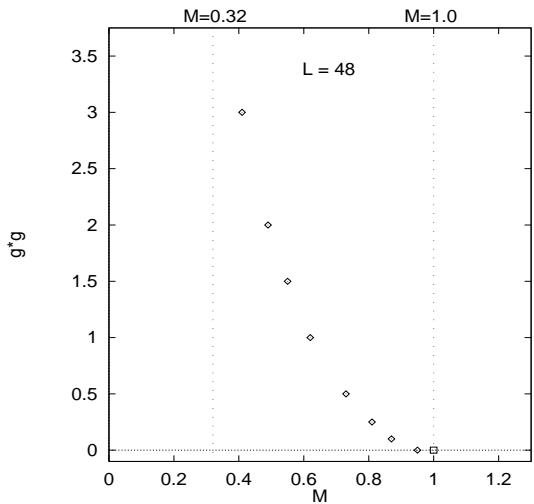

Figure 2. Hartree-Fock phase diagram in $M - g^2$ plane as seen on a finite lattice.

The magnitude of the electric field decreases with gauge coupling and very probably tends to zero in the strong coupling limit. This can be understood as a consequence of the charge shielding. In accordance with Aoki's scenario, $\langle \bar\psi \gamma_5 \psi \rangle$ is also nonzero in the broken region. This expectation however approaches zero (in lattice units) as gauge coupling tends to zero and its behaviour with coupling is complementary to that of the electric field.

To determine the order of the parity-violating phase transition I have calculated the connected correlation functions for electric field and axial charge density in the H-F vacua. The behaviour of these correlators shows that the transition is of first order at finite coupling, approaching a second order endpoint as coupling tends to zero. It should be noted here however, that since the H-F approximation is a mean-field like approach this result should be taken with some care (see also related talk by A. Galante in this volume).

## 3. TWO FLAVOURS

The two flavour case with degenerate masses arises as an obvious generalization of the single flavour Hamiltonian Eq.(1) and the resulting H-F equations are analogous to Eqs.(3,4). I again consider the lattice theory in the space of three parameters $M, M_5$ and $g$, normalizing $K$ and $r$ so that $M_c(0) = 1$. In numerical work, I have allways included a small flavour braking ($\delta = 10^{-3}$) in both $M$ and $M_5$.

While technically there is essentially nothing new here compared to one flavour, the physical behaviour is rather different. Indeed, in terms of surface mode picture, nothing dramatic is expected to happen upon crossing $M_c(g^2)$ in the $M - g^2$ plane. Both flavours will generate their surface modes at that point, each contributing by $\theta = \pi$, thus physically changing nothing. However, going off this plane by turning on $M_5$, interesting things should happen. In particular, there should be phase transitions occuring in the $M - M_5$ plane at the positions where the total value of $\theta$ reaches $\pi$. Indeed, if $\theta$ exceeds $\pi$, it is energetically favourable for the system to reduce the magnitude of the electric field by creating the electron-positron pair. This reverses the sign of the field and induces a transition $\pi \leftrightarrow -\pi$ [3].

In the light of these considerations, I plot in Fig.3 the expectation value of the electric field on the $M_5$-axis at $g^2 = 0.1$. The expected abrupt change of the sign can be nicely seen. Defining the transition point as a position in the $M - M_5$ plane where field switches its sign, I plot in Fig.4 the Hartree-Fock phase diagrams at $g^2 = 0.1$ and $g^2 = 1.2$ observed on a lattice with 32 sites.

The expectations $\langle\overline{\psi}\gamma_5 1\psi\rangle$ and $\langle\overline{\psi}\gamma_5\tau_3\psi\rangle$ are zero in $M - g^2$ plane for $g^2 \leq 1.2$. In fact, there is a very narrow region close to $M_c(g^2)$ where the second of these expectations is nonzero. However, the lattice size dependence of the width of this region indicates its disappearance in the infinite volume limit. For couplings stronger than $g^2 \sim 5$ the Aoki's parity-flavour broken phase appears to be opening up ($\langle\overline{\psi}\gamma_5\tau_3\psi\rangle \neq 0$ despite of the Mermin-Wagner theorem). However, the numerical analysis is not completed yet and I will report on these findings in the detailed account of this work that is in preparation.

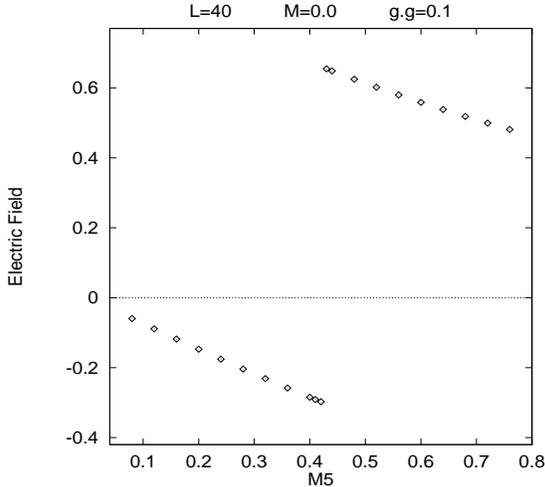

Figure 3. Electric field vs $M_5$ in the model with two flavours at weak coupling.

## 4. CONCLUSIONS

The existence of the line of parity violating phase transitions $M_c(g^2)$ has been established in the Hartree-Fock approximated one flavour Schwinger model on the lattice with Wilson fermions. The physics of the broken phase directly corresponds to $\theta = \pi$ in continuum and also, $\langle\overline{\psi}\gamma_5\psi\rangle \neq 0$ in this phase. I have not found the evidence of phase transitions outside the $M - g^2$ plane ($M_5 \neq 0$), where the physics corresponding to other values of $\theta$ is expected to be recovered.

There is a similar line $M_c(g^2)$ in the mass-degenerate two flavour model. However, at weak couplings (at least for $g^2 \leq 1.2$) the physics is not different in the two regions separated by this line that can be understood as a line consisting of endpoints of the first order ($\theta = \pi \leftrightarrow \theta = -\pi$) phase transitions occuring in the $M - M_5$ plane.

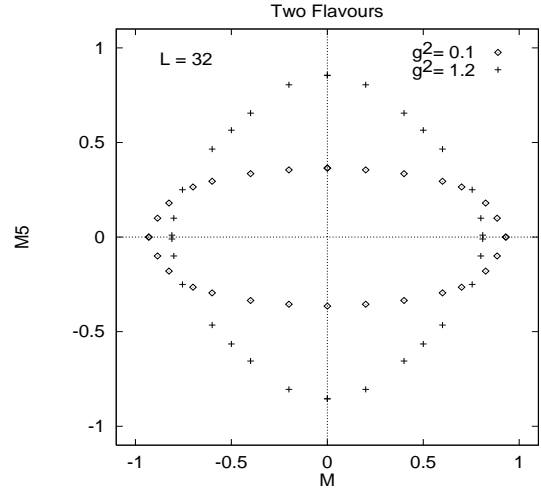

Figure 4. Hartree-Fock phase diagrams in the $M - M_5$ plane as seen on a finite lattice.

## 5. ACKNOWLEDGEMENTS

I would like to thank Mike Creutz, Tony Kennedy, Khalil Bitar and Urs Heller for valuable discussions.